\documentstyle[12pt,epsf]{article}

\def\be{\begin{equation}}
\def\ee{\end{equation}}
\def\bea{\begin{eqnarray}}
\def\eea{\end{eqnarray}}

\def\appendix#1{
  \addtocounter{section}{1}
  \setcounter{equation}{0}
  \renewcommand{\thesection}{\Alph{section}}
 \section*{Appendix \thesection\protect\indent \parbox[t]{11.715cm} {#1}}
  \addcontentsline{toc }{section}{Appendix \thesection\ \ \ #1}}
\def\e{{\rm e}}

\def\cs{(2\pi\alpha')^2}

\def\half{{\frac{1}{2}}}
\def\haf{{\frac{1}{2}}}
\def\tr{{\rm Tr}}
\def\goes{\rightarrow}
\def\goal{\alpha'\rightarrow 0}
\def\exp{{\rm exp}}

\newcommand{\non}{\nonumber \\}

\newcommand{\eg}{{\it e.g.}\ }

\begin{document}

\begin{flushright}
IPM/P-99/043 \\
hep-th/9907146
\end{flushright}

\pagestyle{plain} \vskip .1in

\begin{center}
{\Large {\bf D-Particle Feynman Graphs And Their Amplitudes}\\ }
\vspace{.4cm}

Shahrokh Parvizi$_{a,b}$\hspace{.5 cm}and \hspace{.5 cm}
Amir H. Fatollahi$_{a}$
\vspace{.4 cm}

{\it a)Institute for Studies in Theoretical Physics and Mathematics (IPM),}\\
{\it P.O.Box 19395-5531, Tehran, Iran}\\

\vspace{.2 cm}
{\it b)Shahid Rajaee University, Faculty of Science, Physics
Department,}\\
{\it P.O.Box 16785-163, Lavizan, Tehran, Iran}\\
\vspace{.2 cm}

{\sl parvizi,fath@theory.ipm.ac.ir}
\vspace{.2 cm}
\end{center}
\begin{abstract}
It is argued that quantum traveling of D-particles
 presents the ``joining-splitting'' processes
of field theory Feynman graphs. The amplitudes in $d$ dimensions
can be corresponded with those of a $d+2$ dimensional theory in the
Light-Cone frame. It is shown that this Light-Cone formulation
enables to study processes with arbitrary longitudinal momentum
transfers. It is discussed that a massless sector exists which can
be taken as the low energy limit. By taking the constant relative
distance in the bound-states we find a spectrum for the
intermediatory fields.
\end{abstract}

\vspace{.1cm}
\section{Introduction}
\setcounter{equation}{0}
Perturbative string theory is defined by series expansions based
on the topologies of the string world-sheet, with the weights
proportional to the action of string. The perturbation series can
be presented as sum over different graphs which describe the
``joining-splitting''(JS) processes of strings as events in space-time.
In fact these graphs are those which should be obtained as Feynman graphs
of a {\it string field theory}. In spite of the absence of a
complete string field theory, those graphs which are
produced by small strings ($\goal$ limit) correspond to
well known particle field theories namely
various (super) gravity and gauge theories. So basically one expects
to have a similar interpretation for JS processes of particles and
corresponding particle field theory graphs.

On the other hand, through the developments of understanding of M-theory
as a M(atrix) model \cite{BFSS}, a description of perturbative Light-Cone
string theory was achieved, based on a non-Abelian gauge theory
living on a 2d cylinder which in most times can be interpreted as the
world-sheet of free strings. This description was mentioned
in \cite{Mot,BaSe}, more concreted in \cite{DVV}, and
 has been checked to produce the well known amplitudes
in string theory \cite{Fro1,Fro2}.
Besides in \cite{Wyn} (see also \cite{HV}) it was argued that configurations
with different lengths of strings and their JS processes
are corresponded to various backgrounds of world-sheet (gauge) fields.
By the above picture for Feynman graphs of strings,
it is the purpose of this letter to study similar ones for
D-particles. It is argued that quantum traveling of D-particles
are corresponded to graphs of a certain
particle field theory and their amplitudes in the Light-Cone frame.
So it provides the possibility of defining a particle
field theory by its Feynman graphs, those generated by JS processes of
D-particles.

It is useful to compare our approach to D-particle dynamics and that
of M(atrix) model calculations \cite{BFSS}. In the case of M(atrix) model,
because of large amount of supersymmetry, the D-particle action is trusted
even for large separated D-particles; where it is not expected
the action to be valid \cite{DKPS}.
Though it may be taken an advantage, but lacks
using the action in non-supersymmetric case.
In our approach the dynamics of D-particles is used exactly in the region of
kinematical variables where the action is defined.
It causes that the calculation of D-particle scattering amplitudes can
be performed even in the non-supersymmetric case. The gauge theoretic
description of interactions in bosonic case is supported by the
result of \cite{Fro1} for Light-Cone bosonic strings.

The organization of the article is as follows. In the next section
we review some important aspects of D-particle dynamics.
In section 3 the correspondence between D-particle quantum
mechanics and the Light-Cone
field theory amplitudes is presented. Section 4 is devoted to specification
of the field theory in more details by estimating
the masses of the intermediate states. Section 5 is for conclusion.

\section{On D-Particles}
\setcounter{equation}{0}
D$p$-branes are $p$ dimensional objects which are defined as
(hyper)surfaces which can trap the ends of strings \cite{Po2}.
One of the most interesting aspects of D-brane dynamics appears in their
{\it coincident limit}. In the case of coinciding $N$ D$p$-branes,
their dynamics are captured by dimensional reduction of $U(N)$
gauge theory to $p+1$ dimensions of D$p$-brane world-volume \cite{W,Po2}.

For D-particles $(p=0)$ since only time exists in the world-line
the above dynamics reduces to quantum mechanics of matrices. The
 bosonic action is \cite{BFSS,KPSDF,HALP}
\bea\label{1.1}
S=m_0 \int dt\tr\; \biggl(\half  D_tX_i^2 +\;
\frac{1}{4\cs}\;[X_i,X_j]^2\;\biggl)  ,\;\;\;i=1,...,d,
\eea
where $\frac{1}{2\pi\alpha'}$ and $m_0=(\sqrt{\alpha'}g_s)^{-1}$ are
the string tension and the mass of D-particles respectively, and
$g_s$ is string coupling. Here $D_t=\partial_t-iA_0$ acts as covariant
derivative in the 0+1 dimensional gauge theory.
For $N$ D-particles $X$'s are in adjoint representation of $U(N)$ and have
the usual expansion $X_i=x_{ia}T_a,\;\;a=1,...,N^2$, where
$T_a$'s make basis of $U(N)$ algebra.

In fact this action is the result of the string theory calculation of D-brane
interactions  in the so-called {\it gauge theory limit}, defined by:
\bea\label{1.5}
g_s &\rightarrow& 0,\nonumber\\
l_s &\rightarrow& 0,\nonumber\\
\frac{v}{l_s^2}&=&{\rm fixed},\nonumber\\
\frac{r}{l_s^2}&=&{\rm fixed},
\eea
where $v$ and $r$ are the relative velocities and distances involved in the
kinematics of D-branes and $l_s=\sqrt{\alpha'}$ is the string length.

Firstly let us search for D-particles in the above Lagrangian:\\
For each direction $i$ there are $N^2$ variables instead of $N$ which one
expects for $N$ particles. However there is
a solution for the equations of motion
which restricts the $T_{(a)}$ basis to the $N$ dimensional Cartan
subalgebra. This solution causes vanishing the potential and one
finds the equations of motion for $N$ free particles. In this case the $U(N)$
symmetry is broken to $U(1)^N$ and the interpretation of $N$ remaining
variables as the classical relative positions of $N$ particles is
meaningful. The center of mass of D-particles is represented by the trace
of the $X$ matrices and easily can be seen that the momentum of the
center of mass is conserved.

In the case of unbroken gauge symmetry the gauge transformations
mix the entries of matrices and the interpretation of positions
for D-particles remains obscure \cite{Ba}. But even in this case
the center of mass is meaningful and the ambiguity about positions
only comes back to the relative positions of D-particles.
In unbroken phase the $N^2-N$ non-Cartan elements of matrices have
a stringy interpretation; they govern the dynamics of low lying
oscillations of strings stretched between D-particles.

Let us study the limit $\goal$ more carefully.
Here this limit is analogous to the limit $g_s\goes 0$ in \cite{DVV}.
In this limit to have a finite energy one has to restrict $X$'s to
\bea\label{1.10}
[X_i,X_j]=0,\;\;\;\forall\;i,j,
\eea
and consequently vanishing the potential term in the action.
So the classical action reduces to the action of $N$ free particles
\bea\label{1.15}
S=\int dt \sum_{a=1}^N \half m_0 \dot{x}_a^2.
\eea
However the above observation fails when D-particles arrive
each other. When two D-particles come very near each other two eigenvalues
of $X_i$ matrices will be equal and the corresponding off-diagonal elements
can get non-zero values. This is the same story of gauge symmetry
enhancement. The fluctuations of these off-diagonal elements are
responsible for the interaction in D-particle bound-states.

In the non-coincident case we take simply the action (\ref{1.15}).  It is
based on our expectation for free D-branes in large seperations
\cite{Po2}. For $N$ coincident D-particles one may write the action and
Hamiltonian as,
\bea\label{1.20}
S=\int dt \bigg(\half (Nm_0) \dot{X}^2 +
L_{int}(\hat{x}_a,\dot{\hat{x}}_a)\bigg),\;\;\;a=1,2,..., N^2-1,
\eea
\bea\label{1.25}
H=\frac{P^2}{2(Nm_0)} + H_{int}(\hat{x}_a,\hat{p}_a),
\eea
in which $X$ and $P$ are the position and momentum of the center of mass and
$\hat{x}_a$'s and $\hat{p}_a$'s are the non-Abelian
positions and momenta. 

The dependences of energy eigenvalues and the size of bound states
are notable. By the scalings \cite{KPSDF,DKPS}
\bea
t&\rightarrow& g_s^{-1/3} t,\nonumber\\
A_0&\rightarrow& g_s^{1/3} A_0,\nonumber\\
X&\rightarrow& g_s^{1/3} X,
\eea
one finds the relevant energy and size scales as
\bea
E &\sim& g_s^{1/3}/l_s,\nonumber\\
l_{d+2}&=& g_s^{1/3}l_s.
\eea
The length scale $l_{d+2}$ should be the fundamental length scale of
the covariant $d+2$ dimensional theory whose Light-Cone formulation
is argued to be action (\ref{1.20})
\footnote{In the case $d=9$ this length is known to be the
11-dimensional Planck length.}. In the weak coupling $g_s\to 0$ one
finds $l_{d+2}\ll l_s$ which allows to treat the bound-states of finite
number of D-particles as point like objects in the transverse dimensions
of the Light-Cone frame. Also consequently one finds
$m_0\cdot E \sim l_{d+2}^{-2}$ which shows invariance under boost
transformation of this combination. As we will see the mass of
the intermediate states appear as $\sim m_0\cdot E$.

\section{D-Particle Feynman Graphs}
\setcounter{equation}{0}
Due to momentum conservation, $N$-particle states in Light-Cone
frame have $NP_+$ longitudinal momentum. On the other hand, the longitudinal
momentum of a state plays the role of the Newtonian mass in the transverse
space of Light-Cone frame. So to have $N$-particle (normalizable) states
the existence of marginal bound-states is necessary; those which are
 defined by $E_{binding}=0$. So the energy-mass equivalence
relation for them is read as
\bea
E=M_N=N\cdot m_0
\eea
which in Light-Cone case we have $m_0=P_+$.
The existence of normalizable marginal bound-states in D-particle
case is clarified for supersymmetric case \cite{zero-bound-states}
, and also for bosonic case \cite{gaka}.

Here there are two possibilities to produce specified initial and final
states, the so-called s- and t-channel processes.

\vspace{.5cm}
{\bf s-channel}

\begin{figure}[t]
\begin{center}
\leavevmode
\epsfxsize=80mm
\epsfysize=80mm
\epsfbox{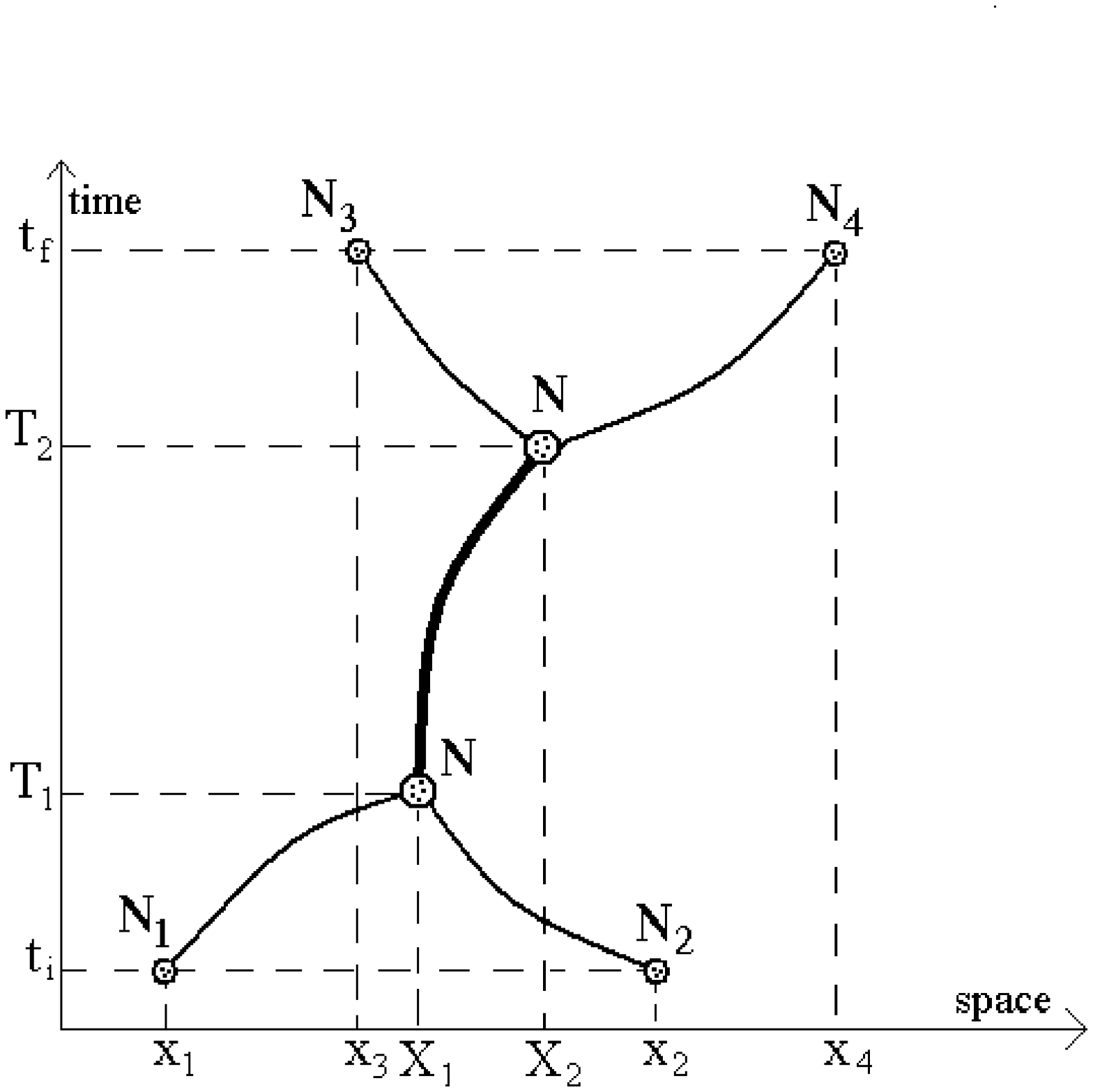}
\caption{{\it s-channel tree path for D-particles, $N=N_1+N_2=N_3+N_4$.}}
\end{center}
\end{figure}

Take the probability amplitude
corresponded to detect four clusters of D-particles,
those with masses $N_1m_0$ and $N_2m_0$ in positions $x_1$ and $x_2$ at time
$t_i$, and masses $N_3m_0$ and $N_4m_0$ in $x_3$ and $x_4$ at time $t_f$,
presented by path integral as
\bea\label{2.60}
\langle  x_3, x_4;t_f | x_1,x_2;t_i \rangle =\int \e^{-S},
\eea
and supplied by the Newtonian mass conservation
\bea
N_1+N_2=N_3+N_4=N.
\eea
In the limit $\goal$ one writes (Fig.1)
\footnote{Here as in field theory we
have dropped the dis-connected graphs. This work can be done by subtracting
the contribution of paths corresponded to free moving of D-particles.}
\bea\label{2.65}
\langle x_3, x_4;t_f | x_1,x_2;t_i \rangle &=&\left[\int \e^{-S}\right]_{\goal}\non
         &=&\int_{t_i}^{t_f}dT_1 dT_2
         \int_{-\infty}^{\infty} d^dX_1 d^dX_2  \nonumber\\
 &\times& \bigg(K_{N_1m_0}(X_1,T_1;x_1,t_i)
 K_{N_2m_0}(X_1,T_1;x_2,t_i)\bigg)
\nonumber\\
&\times&
\bigg(K_{Nm_0}(X_2,T_2;X_1,T_1)
K_{int}(\hat x_{2a}=0,T_2;
\hat x_{1a}=0,T_1)\bigg)
\nonumber\\
&\times&
 \bigg(K_{N_3m_0}(x_3,t_f;X_2,T_2)
 K_{N_4m_0}(x_4,t_f;X_2,T_2)\bigg),
\eea
which $K_{m}(y_2,t_2;y_1,t_1)$ is the non-relativistic propagator
of a free particle with mass $m$ between $(y_1,t_1)$ and $(y_2,t_2)$
, and  $K_{int}$ is the corresponding propagator for
the non-Abelian path integrations over $L_{int}(\hat{x}_a,\dot{\hat{x}}_a)$.
 In the above $\int dT_1 dT_2 dX_1 dX_2$ is for a summation over
different JS times and points. It is assumed that the group
$U(1)\otimes SU(N_1)\otimes SU(N_2)$ for the position matrices is
enhanced as $U(1)\otimes SU(N)$ at time $T_1$ and again is broken to
$U(1)\otimes SU(N_3)\otimes SU(N_4)$ at time $T_2$.

We use the representations in $d$ dimensions
\bea\label{2.40}
K_{m}(y_2,t_2;y_1,t_1)&=&\theta(t_2-t_1)
\frac{1}{(2\pi)^d} \int d^dp
\;\exp\bigg(ip\cdot(y_2-y_1)-\frac{ip^2(t_2-t_1)}{2m}\bigg),
\\
K_{int}(\hat x_{2a},T_2;\hat x_{1a},T_1)
&=&\int d\hat{x}_a \e^{-S_{int}[T_2,T_1]}=
\sum_{n} \langle \hat{x}_{2a}|n\rangle \langle n|\hat{x}_{1a}\rangle
\e^{-iE_n(T_2-T_1)},
\eea
where $E_n$'s are the eigenvalues of $H_{int}$ of (\ref{1.25}) and
$\theta(t_2-t_1)$ is the step function.

It is more convenient to translate every thing to the momentum space by
 ($E_k=\frac{p_k^2}{2N_km_0}$, $k=1,2,3,4$)
\bea\label{2.75}
\langle  p_3,p_4,E_3,E_4;t_f |p_1, p_2,E_1,E_2;t_i  \rangle &\sim&
\e^{i(E_3+E_4)t_f-i(E_1+E_2)t_i}\int\prod_{k=1}^4 d^dx_k \times
\nonumber\\
&~&  \e^{i(p_1x_1+p_2x_2-p_3x_3-p_4x_4)}
\langle  x_3,x_4;t_f | x_1, x_2;t_i \rangle,
\eea
and doing all integrations one finds (for $t_i=-\infty$ and $t_f=\infty$)
\footnote{We recall
$$
\int_0^\infty \e^{i\alpha\xi}=\lim_{\epsilon\rightarrow 0^+}
\frac{i}{\alpha+i\epsilon}.
$$}
\bea\label{2.85}
\langle  p_3,p_4,E_3,E_4;\infty |p_1, p_2,E_1,E_2;-\infty  \rangle &\sim&
\delta^{(d)}(p_1+p_2-p_3-p_4)\delta(E_1+E_2-E_3-E_4)
\nonumber\\
&~&\times \lim_{\epsilon \rightarrow 0^+}
\sum_n\frac{i\langle \hat{x}_{2a}=0|n\rangle \langle n|\hat{x}_{1a}=0\rangle}
{E-\frac{q^2}{2(Nm_0)}-E_n+i\epsilon},
\eea
where $\vec q=\vec p_1 +\vec p_2$ and $E=E_1+E_2$.
 Now recalling the energy-momentum relation in the Light-Cone frame
for a particle with mass $M$
$$
E\equiv q_-=\frac{{\vec{q}}^2 +M^2}{2q_+},
$$
one sees that the fraction in the sum of (\ref{2.85}) is
the the Light-Cone propagator of a particle \cite{Suss} by identifications
\bea\label{2.90}
q_+&=&Nm_0,\nonumber\\
M_n^2&=&2Nm_0 E_n.
\eea
The first relation of (\ref{2.90}) learns us that each D-particle has
Light-Cone momentum equal to $m_0$; this is because to have the Light-Cone
momentum $Nm_0$ for $N$ D-particles in the time interval $[T_1,T_2]$
(Fig.1). So (\ref{2.85}) is the same expression which one writes
(in momentum space) as tree diagram contribution to 4-point function of
a field theory but in the Light-Cone frame \cite{Suss}, with exchanging
masses as $M_n$'s. It is noted that $M_n^2\sim 1/l_{d+2}^2$.

In a covariant form one writes for (\ref{2.85})
\bea\label{2.95}
\langle  p_3^\mu,p_4^\mu;\infty |p_1^\mu, p_2^\mu;-\infty  \rangle &\sim&
\delta(\vec p_1+\vec p_2-\vec p_3-\vec p_4)
\delta(p_{-1}+p_{-2}-p_{-3}-p_{-4})
\nonumber\\
&~&\times \lim_{\epsilon \rightarrow 0^+}
\sum_n
\frac{i2Nm_0\langle \hat{x}_{2a}=0|n\rangle \langle n|\hat{x}_{1a}=0\rangle}
{ q_\mu q^\mu - M_n^2+i\epsilon},
\eea
where $q^\mu=p_1^\mu+p_2^\mu$ and for any vector $V$ we have
$$
V^\mu\equiv (\frac{V_++V_-}{\sqrt 2},\vec V,\frac{V_+-V_-}{\sqrt 2}),
\;\mu=0,1,2,...,d,d+1.
$$
 So
$$
p_{-k}\equiv \frac{ \vec p_k^2}{2N_km_0}, \;\;\; p_{+k}\equiv N_km_0,
\;\;k=1,2,3,4.
$$

\vspace{.5cm}
{\bf t-channel}

The above considered amplitude for detecting four clusters of D-particles
also can be studied in t-channel as (Fig.2)
\bea\label{2.105}
\langle x_3, x_4;t_f | x_1,x_2;t_i\rangle &=&
\left[\int \e^{-S}\right]_{\goal}\non
         &=&\int_{t_i}^{t_f}dT_1 dT_2
         \int_{-\infty}^{\infty} d^dX_1 d^dX_2  \nonumber\\
 &\times& \bigg(K_{N_1m_0}(X_1,T_1;x_1,t_i)
 K_{N_3m_0}(x_3,t_f;X_1,T_1)\bigg)
\nonumber\\
&\times&
\bigg(K_{N'm_0}(X_1,T_1;X_2,T_2)
K_{int}(\hat x_{2a}=0,T_1;
\hat x_{1a}=0,T_2)\bigg)
\nonumber\\
&\times&
 \bigg(K_{N_2m_0}(X_2,T_2;x_2,t_i)
 K_{N_4m_0}(x_4,t_f;X_2,T_2)\bigg),
\eea
with the definitions mentioned earlier. In this channel as is specified in
Fig.2, we have $T_1>T_2$. This is because of
the unique direction of time in Light-Cone frame; see (\ref{2.40}).
So for every propagation of particles, \eg intermediate states,
propagations are from smaller time to bigger time.

\begin{figure}[t]
\begin{center}
\leavevmode
\epsfxsize=80mm
\epsfysize=80mm
\epsfbox{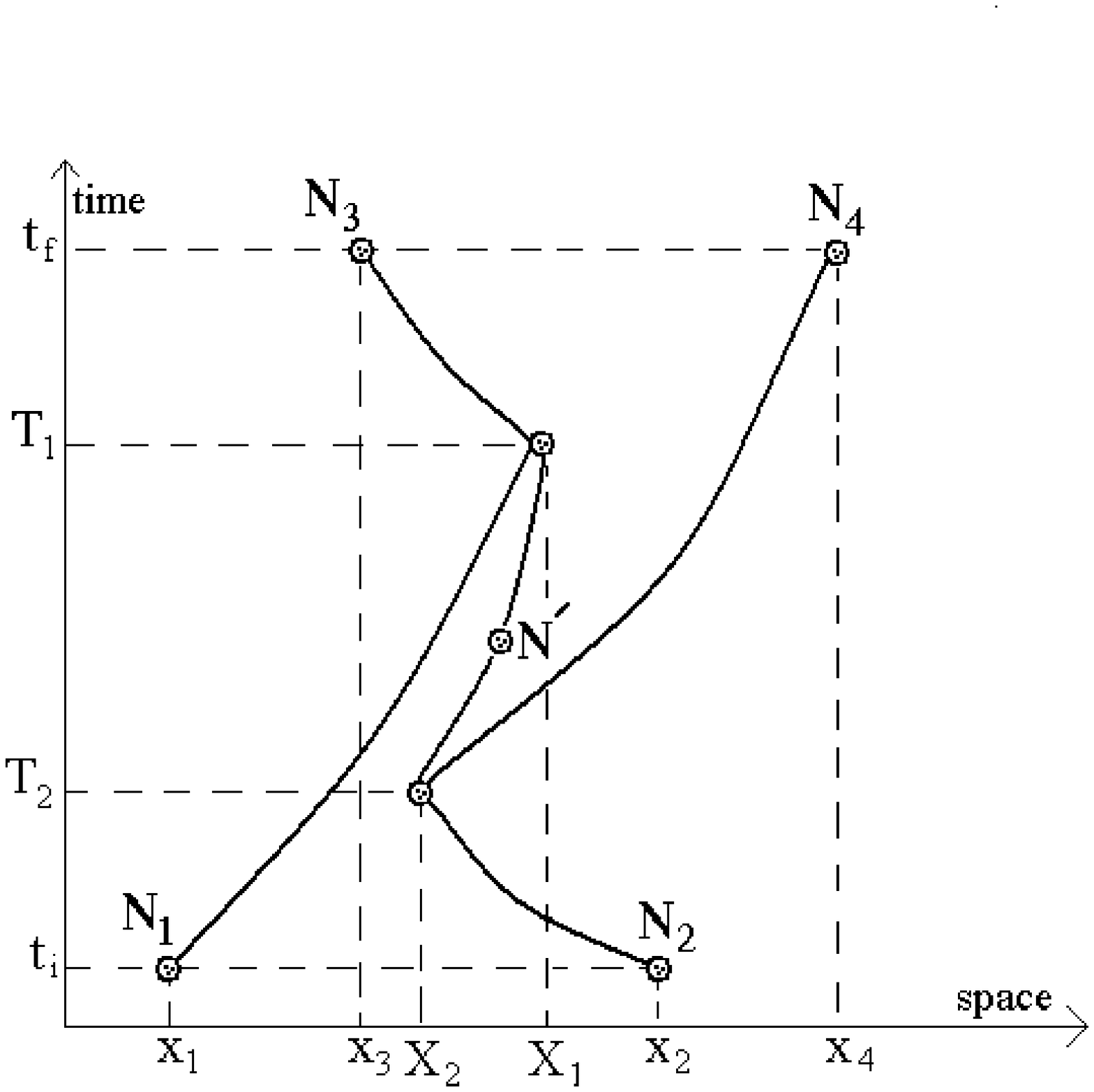}
\caption{{\it t-channel tree path for D-particles, $N'=N_2-N_4=N_3-N_1$.}}
\end{center}
\end{figure}

Again by going to the momentum space
and doing all integrations one finds (for $t_i=-\infty$ and $t_f=\infty$)
\bea\label{2.110}
\langle  p_3,p_4,E_3,E_4;\infty |p_1, p_2,E_1,E_2;-\infty  \rangle &\sim&
\delta^{(d)}(p_1+p_2-p_3-p_4)\delta(E_1+E_2-E_3-E_4)
\nonumber\\
&~&\times \lim_{\epsilon \rightarrow 0^+}
\sum_n\frac{i\langle \hat{x}_{2a}=0|n\rangle \langle n|\hat{x}_{1a}=0\rangle}
{(E_3-E_1)-\frac{q'^2}{2(N'm_0)}-E_n+i\epsilon},
\eea
where $\vec{{q'}}=\vec p_3 -\vec p_1$ and $N'=N_3-N_1$.
Again we have Light-Cone propagator of a particle \cite{Suss}
by identifications
\bea\label{2.115}
q'_+&=&N'm_0,\nonumber\\
M_n^2&=&2N'm_0 E_n.
\eea

In a covariant form one writes for (\ref{2.110})
\bea\label{2.120}
\langle  p_3^\mu,p_4^\mu;\infty |p_1^\mu, p_2^\mu;-\infty  \rangle &\sim&
\delta(\vec p_1+\vec p_2-\vec p_3-\vec p_4)
\delta(p_{-1}+p_{-2}-p_{-3}-p_{-4})
\nonumber\\
&~&\times \lim_{\epsilon \rightarrow 0^+}
\sum_n
\frac{i2N'm_0\langle \hat{x}_{2a}=0|n\rangle \langle n|\hat{x}_{1a}=0\rangle}
{ q'_\mu q'^\mu - M_n^2+i\epsilon},
\eea
where $q'^\mu=p_3^\mu-p_1^\mu$.

\section{Field Theory Of Feynman Graphs}
\setcounter{equation}{0}
The question of ``what field theory?'' is answered only by knowing the
full set of eigenvalues and eigenvectors of the Hamiltonian $H_{int}$
defined by,
$$
H_{int}|n\rangle =E_n |n\rangle .
$$
The knowledge about the spectrum of the interested Hamiltonian
is restricted. As is noted earlier the existence of zero-energy
bound-states is discussed in \cite{zero-bound-states,gaka}
\footnote{In the bosonic case these states are formed by
identified D-particles, with the corresponding eigen-function
to be constant in internal space \cite{gaka}.}.
So, as the same of string theory, we have massless states
which produce the low energy limit of the $d+2$
dimensional theory.

In the following we discuss another approach, the constant background.


\subsection{Constant background}
When two D-particles come very near each other two eigenvalues
of $X_i$ matrices will become approximately equal and this
makes the possibility that the
corresponding off-diagonal elements take non-zero values.

In the coincident limit the dynamics is complicated.
The relative matrix position may be taken as:
\bea
X_i= \left( \matrix{x_i & Y_i \cr
Y_i^* & -x_i }\right),
\eea
where $Y^*$ is the complex conjugate of $Y$. By inserting this matrix in
the Lagrangian one obtains:
\bea
S=\int dt\haf \bigg( (2m_0) \dot{X_i}^2 +  m_0 \dot{Y}_i\dot{Y}_i^*
-  m_0 \frac{1}{4\cs} (1-\cos^2\theta)
x_j^2 Y_iY_i^*+m_0 \dot{x}_j^2+O(Y^3)\bigg),
\eea
with $X$ for the center of mass and $\theta$ is the angle between $\vec{x}$
and the complex vector $\vec{Y}$.
As is apparent in the limit $\goal$ which is in our direct interest,
the $x$ element can not get large values and have a small range of
variation. In high-tension approximation of strings, one takes the
relative distance of D-particles constant and of order of bound-state size
$x\sim g_s^{1/3}l_s$, as was mentioned in Sec.2. So one writes:
\bea
S=\int dt \bigg(\haf (2m_0) \dot{X_i}^2 +
\haf m_0 \dot{Y}_\perp\cdot\dot{Y}_\perp^*
- \haf m_0 \frac{k^2g_s^{2/3}}{\alpha'}
Y_\perp\cdot Y_\perp^* +\cdot\cdot\cdot
\bigg),
\eea
where in the above $k$ is an $\alpha'$ independent numerical factor, and
$Y_\perp$ is the perpendicular part of the $\vec{Y}$ to the relative distance
$\vec{x}$. The parallel part of $\vec{Y}$ behaves as a free part.
In $d+1$ dimensions of space-time the dimension of $Y_\perp$ is
$d-1$ which shows that we are encountered with $2\times(d-1)$
harmonic oscillators because, $Y$ is a complex variable
\footnote{This is the same number of harmonic oscillators which
appear in one-loop calculations \cite{9902}}. These harmonic oscillators
are corresponded to vibrations of oriented open strings stretched between
D-particles.

By translating all the above to the momentum space
and doing the integrals it is found
\bea
\langle p_3,p_4,E_3,E_4;t_f |p_1, p_2,E_1,E_2;t_i\rangle &\sim&
\delta^{(d)}(p_1+p_2-p_3-p_4)
\nonumber\\
&~&\int_{t_i}^{t_f} dT_1 dT_2 \theta(T_2-T_1)
\times\exp(\frac{-i(p_1^2+p_2^2)T_1}{2m_0})
\nonumber\\
&~&\exp(\frac{-iq^2(T_2-T_1)}{4m_0})
\exp(\frac{i(p_3^2+p_4^2)T_2}{2m_0})
\nonumber\\
&~&
K_{oscillator}(Y_\perp=0,T_2;Y_\perp=0,T_1)
\eea
where $\vec{q}=\vec{p}_1+\vec{p}_2=\vec{p}_3+\vec{p}_4$, and
$$
K_{oscillator}(Y_\perp=0,T_2;Y_\perp=0,T_1)=
\theta(T_2-T_1)\bigg(\frac{m_0\omega}{2\pi i
\sin[\omega(T_2-T_1)]}\bigg)^{d-1},
$$
which $\omega$ is the harmonic oscillator frequency here to be
$kg_s^{1/3}/l_s$.
Since $Y_\perp$'s are complex numbers the
power for the harmonic propagator is twice of $\frac{d-1}{2}$.

To have a real scattering process one assumes
$$
t_i\rightarrow -\infty,\;\;\;t_f\rightarrow \infty.
$$
We put $T\equiv T_2-T_1$ which has the range $0\leq T \leq \infty $.
The integrals yield
\bea
\langle p_3,p_4;\infty |p_1, p_2;-\infty\rangle &\sim&
\delta^{(d)}(p_1+p_2-p_3-p_4)
\delta(\frac{p_1^2}{2m_0}+\frac{p_2^2}{2m_0}-
\frac{p_3^2}{2m_0}-\frac{p_4^2}{2m_0})
\nonumber\\
&~&
\int_0^\infty dT\; \e^{\frac{-iT}{4m_0}(q^2-2(p_1^2+p_2^2))}
\bigg(\frac{m_0\omega}{\sin (\omega T)}\bigg)^{d-1},
\eea
By recalling the energy-momentum relation in the Light-Cone gauge
one has:
$$
2(p_1^2+p_2^2)-q^2= 2(2m_0)(\frac{p_1^2+p_2^2}{2m_0})-q^2=
2q_+q_--q^2=q_\mu q^\mu\equiv q_\mu^2.
$$
So it is found:
\bea
\langle p_3,p_4,E_3,E_4;\infty |p_1,p_2,E_1,E_2;-\infty\rangle
&\sim&
\delta^{(d)}(p_1+p_2-p_3-p_4)\delta(E_1+E_2-E_3-E_4)
\nonumber\\
&~&\int_0^\infty dT \e^{\frac{-q_\mu^2}{4m_0}T}
\bigg(\frac{m_0\omega}{\sin (\omega T)}\bigg)^{d-1}.
\eea
We perform a cut-off for $T$ in small values as
$0 < \epsilon\leq T \leq \infty $,
with $\epsilon$ be small. By changing the integral
variables as $\e^{-2\omega T}=\eta$ we have
\bea\label{amp}
\langle p_3^\mu,p_4^\mu;\infty |p_1^\mu,p_2^\mu;,-\infty\rangle &\sim&
\delta^{(d)}(p_1+p_2-p_3-p_4)\delta(p_{-1}+p_{-2}-p_{-3}-p_{-4})
\nonumber\\
&~&\frac{(m_0\omega)^{d-1}}{2\omega}
\int_0^x d\eta\; \eta^{\frac{-q_\mu^2}{8m_0\omega}+\frac{d-3}{2}}
(1-\eta)^{-d+1},\nonumber\\
&\sim&
\delta^{(d)}(p_1+p_2-p_3-p_4) \delta(p_{-1}+p_{-2}-p_{-3}-p_{-4})
\nonumber\\&~&
\frac{(m_0\omega)^{d-1}}{2\omega}
B_x(\frac{-q_\mu^2}{8m_0\omega}+\frac{d-1}{2},-d+2)
\eea
with $1\sim x=\e^{-2\omega\epsilon}$ and $B_x$ is the Incomplete
Beta function. It is recalled that the $m_0\omega$ is $1/l_{d+2}^2$.
The longitudinal momentum conservation trivially is satisfied.

\vspace{.5cm}
{\bf Polology}

Equivalently one may use the other representation of $K_{oscillator}$
as
\bea
K_{oscillator}(Y_\perp=0,T_2;Y_\perp=0,T_1)=
\sum_{n} \langle0|n\rangle\langle n|0\rangle \e^{-iE_n(T_2-T_1)},
\eea
with $E_n$'s as the known $H_{oscillator}$ eigenvalues. By this
representation one finds the pole expansion :
\bea
\langle p_3^\mu,p_4^\mu;\infty |p_1^\mu, p_2^\mu;-\infty  \rangle&\sim&
\delta^{(d)}(p_1+p_2-p_3-p_4)
\delta(p_{-1}+p_{-2}-p_{-3}-p_{-4})
\nonumber\\
&~&\times \lim_{\epsilon \rightarrow 0^+}
\sum_n C_n\; \frac{i4m_0}{ q_\mu q^\mu - M_n^2+i\epsilon}.
\eea

This pole expansion also can be derived by extracting the poles
of the amplitude (\ref{amp}) with the condition
\bea
\frac{-q_\mu^2}{8m_0\omega}+\frac{d-1}{2}=-n, \;\;
{\rm with}\; n\; {\rm as\; a\; positive\; integer}
\eea
or
\bea
M_n^2=\frac{8k(n+\frac{d-1}{2})}{l_{d+2}^2}.
\eea

\section{Conclusion}
\setcounter{equation}{0}
It is argued in this work that the quantum mechanics of D-particles in $d$
dimensions can be corresponded to Light-Cone formulation of a $d+2$ dimensional
field theory. The length scale of the $d+2$ dimensional theory is
 $g_s^{1/3}l_s$. Also it is shown that this Light-Cone formulation enables
to study processes with arbitrary longitudinal momentum transfers.
It is discussed that a massless sector exists which can be taken as the low
energy limit of the $d+2$ dimensional theory. By taking the constant
relative distances in the bound-states we found a spectrum for the
intermediatory fields.

In the supersymmetric case a candidate for the $d+2$ dimensional theory
can be guessed by M(atrix) theory approach to M-theory \cite{BFSS}.
M-theory and type IIA string theory both have
supergravities in their low energy limit in 11 and 10 dimensions
respectively. There is a gauge theoretic
description of stringy ``Feynman Graphs'' \cite{Mot,BaSe,DVV}.
Since the 10 dimensional (IIA) supergravity is corresponded
to stringy ``Feynman Graphs'' in the limit $\goal$,
so one may expect to have a similar relation between
D-particles ``Feynman Graphs'' and 11 dimensional supergravity.


\end{document}